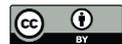

# New regional stratigraphic insights from a 3D geological model of the Nasia sub-basin, Ghana, developed for hydrogeological purposes and based on reprocessed B-field data originally collected for mineral exploration

Elikplim Abla Dzikunoo[1], Giulio Vignoli[2,3], Flemming Jørgensen[4], Sandow Mark Yidana[1], and Bruce Banoeng-Yakubo[1]

[1]Department of Earth Science, University of Ghana, Accra, Ghana
[2]DICAAR, University of Cagliari, Cagliari, Italy
[3]GRUK, Geological Survey of Denmark and Greenland (GEUS), Aarhus, Denmark
[4]Central Denmark Region, Viborg, Denmark

**Correspondence:** Sandow Mark Yidana (yidanas117@gmail.com) and Elikplim Abla Dzikunoo (eadzikunoo@gmail.com)

**Abstract.** Reprocessing of regional-scale airborne electromagnetic data is used to build a 3D geological model of the Nasia sub-basin, northern Ghana. The resulting 3D geological model consistently integrates all the prior pieces of information brought by electromagnetic data, lithologic logs, ground-based geophysical surveys, and geological knowledge of the terrain. The geo-modeling process is aimed at defining the lithostratigraphy of the area, chiefly to improve the stratigraphic definition of the area, and for hydrogeological purposes. The airborne electromagnetic measurements, consisting of GEOTEM B-field data, were originally collected for mineral exploration purposes. Thus, those B-field data had to be (re)processed and properly inverted as the original survey and data handling were designed for the detection of potential mineral targets and not for detailed geological mapping. These new geophysical inversion results, compared with the original conductivity–depth images, provided a significantly different picture of the subsurface. The new geophysical model led to new interpretations of the geological settings and to the construction of a comprehensive 3D geo-model of the basin. In this respect, the evidence of a hitherto unexposed system of paleovalleys could be inferred from the airborne data. The stratigraphic position of these paleovalleys suggests a distinctly different glaciation history from the known Marinoan events, commonly associated with the Kodjari formation of the Voltaian sedimentary basin. Indeed, the presence of the paleovalleys within the Panabako may be correlated with mountain glaciation within the Sturtian age, though no unequivocal glaciogenic strata have yet been identified. Pre-Marinoan glaciation is recorded in rocks of the Wassangara group of the Taoudéni Basin. The combination of the Marinoan and, possibly, Sturtian glaciation episodes, both of the Cryogenian period, can be an indication of a Neoproterozoic Snowball Earth. Hence, the occurrence of those geological features not only has important socioeconomic consequences – as the paleovalleys can act as reservoirs for groundwater – but also from a scientific point of view, they could be extremely relevant as their presence would require a revision of the present stratigraphy of the area.

## 1 Introduction

The present research demonstrates the effectiveness of the reprocessing and proper inversion of existing airborne electromagnetic (AEM) data – more specifically, GEOTEM B-field measurements – for the data-driven inference of subsurface geology. More specifically, the AEM results are employed to develop a 3D geological model for subsequent hydrogeological conceptualization and scenario simulations of groundwater recharge and abstraction (under different environmen-tal and anthropic stresses) in the partially metamorphosed sedimentary Nasia Basin (a sub-catchment within the White Volta Basin in northern Ghana). In fact, the overall objective of the research is to develop a decision-support tool for understanding groundwater occurrence to facilitate the efficient development and optimization of the water resources in the basin within the framework of the GhanAqua project.

The use of groundwater resources for crop irrigation offers an opportunity for a buffer against the unremitting impacts of climate change in northern Ghana, where peasant farming is the mainstay of livelihood. The development of groundwater resources to support irrigation endeavors is particularly important because of erratic rainfall patterns during the rainy season and high temperatures and evapotranspiration rates in the dry season, which render surface water resources unsustainable for irrigation water (Eguavoen, 2008). Erratic rainfall patterns in the region in recent times have affected crop production and the sustainable livelihoods of communities. Hence, improved access to groundwater resources for year-round irrigation would boost agricultural development and offer increased employment possibilities in the area. However, over the years, access to sustainable groundwater resources has been hampered by the lack of sufficient knowledge of the local and structural geological setting. Such knowledge is crucial to the understanding of the hydrogeology and groundwater storage conditions and would be crucial for sustainable resource development.

Generally, the difficulty in defining and effectively characterizing subsurface geological conditions in an area such as the Voltaian sedimentary basin hinges on the unavailability of reliable data (e.g., lithological logs of deep boreholes) and the limitations inherent in conventional ground-based geophysical techniques (e.g., poor spatial coverage and insufficient density). So, a multi-scale holistic approach, integrating the airborne geophysical insights with all the available lithological borehole logs and ground-based geological investigations, is shown to be essential for the development of an effective and coherent geological model to eventually be used for hydrogeological assessments.

Three-dimensional (3D) geological modeling based on specifically collected AEM data for hydrogeological applications is, in general, not new (Jørgensen et al., 2013, 2015; Høyer et al., 2015; Oldenborger et al., 2014), but as far as we are concerned, it has never been done before using B-field measurements. Additionally, geological modeling for hydrogeological application is novel in the West African subregion, even though the region has a rich database of preexisting AEM data from former mineral exploration surveys. Hence, the application of the presented workflow for the inversion of AEM data can potentially be extended to many areas in this part of the African continent and, in general, everywhere preexisting AEM data are available. This can help avoid the costs connected with the airborne data collection, which is often considered affordable for mineral exploration but prohibitive for groundwater mapping.

Besides the (re)use of the abovementioned geophysical datasets and in an attempt to address the main socioeconomic issues connected with an effective hydrogeological characterization of the Nasia Basin, the present research brings some contributions to the geological and stratigraphic knowledge of the Volta Basin. The lithostratigraphy of the sedimentary infill of the Volta Basin is still disputed (Blay, 1983; Affaton, 1990; Carney et al., 2008, 2010). However, there is an overall consensus on a subdivision into three groups (Affaton, 1990; Affaton et al., 1980, 1991; Bertrand-Sarfati et al., 1991): Bombouaka, Oti (or Pendjari), and Obosum. In this research, we provide possible insights on the delineation of the interfaces between the formations characterizing the Nasia portion of the Volta Basin, i.e., Bombouaka and Oti.

## 2 Data and methods

### 2.1 The study area

The approximately 5300 km$^2$ area of the Nasia Basin is found in the northern region of Ghana, within the Guinea Savannah belt. It is associated with an average annual rainfall of 1000–1300 mm, which peaks between late August and early September. Torrential rains within this peak season create serious drainage problems as the infiltration rates are low due to the largely impervious nature of the various lithologies, creating high amounts of runoff, in turn leading to high levels of erosion and posing significant constraints on agriculture (Siaw, 2001).

The area is characterized by relatively low relief in the south and a few areas of high elevation associated with the Gambaga escarpment to the north. The basin drains a left bank tributary of the White Volta, the Nasia River (Fig. 1a), and is underlain by sedimentary rocks of the Bombouaka and Oti–Pendjari groups of the Neoproterozoic Voltaian supergroup; it is comprised predominantly of variations of sandstones, siltstones, and mudstones (Carney et al., 2010). Detailed descriptions of the geologic units can be found in Carney et al. (2010) and also in Jordan et al. (2009).

### 2.2 Data and modeling requirements

A 3D geo-model of an area is a synthesis of all relevant geologic information available; during the construction process, it is essential to integrate and merge multiple data sources and scales in order to appropriately represent the different aspects of complex geologic systems (e.g., Dzikunoo et al., 2018; Rapiti et al., 2018; Jørgensen et al., 2017, 2015; Vignoli et al., 2017, 2012; Høyer et al., 2017). In this regard, the diverse kinds of data used in this study consist of (i) AEM data (namely, GEOTEM B-field), (ii) borehole lithological and geophysical logs, and (iii) preexisting outcrop analyses and geological information.

An important underlying consideration for the construction of a lithostratigraphic model is the definition of a con-ceptual model initially developed from the prior knowledge of the terrain (Fig. 1b). Interpretations from geophysical signatures are then tied into the conceptual model, followed by the development of a model framework with interpretation points and subsequently populated with voxels, each characterized by homogeneous attributes. Clearly, any piece of information brought in this specific case from the geophysics can and must be used, via a confirmation–rejection process, to refine the initial geological hypotheses (Tarantola, 2006).

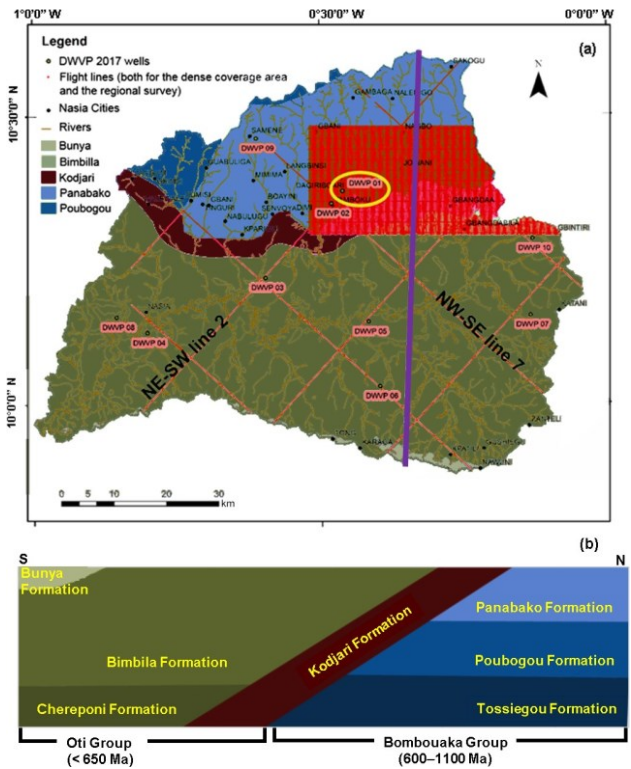

**Figure 1. (a)** Geologic map of the Nasia sub-basin after Carney et al. (2010); the rectangular red area on the top right corner shows the locations of the flight lines of the dense coverage AEM survey (line spacing of 200 m; Fig. 9). The flight lines of the regional survey (20 km by 20 km regular spacing) are shown as red lines (two of them are indicated as NE–SW line 2 and NW–SE line 7). **(b)** Conceptual model of the geology along a cross-sectional N–S line within the study area – solid violet line in panel **(a)**. The age determinations for the Oti and Bombouaka groups are those reported in Kalsbeek et al. (2008).

### Airborne electromagnetic data (AEM)

A fixed-wing Casa 212 aircraft, equipped with a 20-channel GEOTEM multi-coil system, was used to acquire the time-domain electromagnetic data (Fugro Airborne Surveys, 2009a, b) with a line spacing of 20 km (flown at 042–132° along and across the general geologic strike lines within the Volta Basin) and a much denser line spacing of 200 m (flown at 000–180°). The locations of the flight lines of both surveys (dense and regional) are shown as red lines in Fig. 1a (the 200 m spacing makes the dense survey appear as a red rectangle). The GEOTEM surveys were performed under the auspices of the European Union Mining Sector Support Programme 2005 to 2008 and were designed for mineral exploration.

Within the present study, the original B-field data have been reprocessed and inverted; this is to ensure the preservation of all the corrections applied to the raw data by Fugro and, contextually, to have the opportunity to consistently compare the new outcomes with the conductivity–depth images (CDIs) provided by the survey company as final deliverables (Fugro Airborne Surveys, 2009a, b). This comparison was necessary in order to estimate what could be gained by going through a complete reprocessing and inversion in terms of the reliability and accuracy of the subsequent (hydro)geological model(s). Since the data acquisition was originally focused on mineral targeting, the specifications of the survey and the choice of CDIs as deliverables were intended to optimize the detection, even at depth, of large conductivity contrast targets (typical for mineral exploration) with a potentially high lateral resolution. Conversely, for geological mapping purposes, the capability to retrieve even low-contrast conductivity features via proper inversion strategies and, at the same time, to reproduce the spatial coherence of the geological features is crucial. Therefore, it was important to double-check the effectiveness of the new inversion approaches and of the dedicated preliminary data conditioning.

In addition to the advantages discussed in Smith and Annan (1998), the choice of B-field has some further benefits in terms of the signal-to-noise ratio as the B-field is associated with data integration over time that can act as some sort of stacking in time. Clearly, the data stacking can (and should) also be performed in the other "direction", that is, spatially along the line of flight. In the workflow implemented in this research, a moving window with a width variable depending on the considered time gate has been used in a fashion similar to the one detailed, for example, in Auken et al. (2009) and Vignoli et al. (2015a) (but here the stacking window width is frequency-dependent). This strategy allows for the use of (i) a narrower time window at the early gates and (ii) a wider window at late gates where the signal has, in any case, a larger spatial footprint. By doing so, we can obtain the maximum spatial resolution at the near surface (where the signal is stronger) and, contextually, improve as much as possible the signal-to-noise ratio at depth (where the physics of the method naturally average the information). In particular, the size of the window utilized for the $Z$ component of the B-field is linearly increasing from 7.911 s for the first gate (4.505 ms) to 19.876 s for the 11th gate (11.563 ms) and remains equal to 20 s for the last four gates; concerning the $X$ component, only 12 gates have been used, but with the same stacking window settings. An example of the data resulting from the application of this moving window strategy is shown in Fig. 2. In practice, these specific settings for the moving window have been selected through a visual trial-and-adjustment procedure aimed at removing the suspicious oscillations of the signal without laterally smoothing the data too much.

With respect to the inversion – as the receiver of the GEOTEM system is located in a towed bird – the altitude, pitch, and roll of the device were part of the inversion parameters, and they were reconstructed by using the $Z$ and $X$ components of the B-field measurements and by enforcing a lateral continuity between nearby measurement locations. A similar approach has also been used for the main model parameter we inverted for: the electrical resistivity. On the other hand, the thicknesses of the 30-layer parameterization have been kept fixed and were not involved in the inversion. So, in the framework of a pseudo-3D inversion based on a 1D forward modeling, the resistivity of a specific discretizing layer was coupled to the resistivity values at the corresponding depths in the adjacent 1D soundings. Thus, for the data collected with a 200 m line spacing, we applied the so-called spatially constrained inversion (SCI; Viezzoli et al., 2008), while for the less dense acquisition data (20 km line spacing), its 2D version (laterally constrained inversion – LCI) was used instead. The final results clearly depended on the specific choices of the inversion settings (e.g., the relative weight of the regularizing term connecting the resistivities of layers of different soundings) and the choices of the inversion strategy (e.g., sharp versus smooth regularization – Auken et al., 2014; Ley-Cooper et al., 2014; Vignoli et al., 2015b, 2017). In order to select the most effective regularization capable of retrieving a resistivity distribution compatible with (i) the observations within the noise level and (ii) the most reasonable geological expectation, we adopted an iterative geophysics–geology approach, characterized by a close interaction between geologists and geophysicist (Fig. 3). The basic rationale is that the geological interpretation already starts at the geophysical processing stages. The inversion of B-field curves (inevitably characterized by a finite number of gates and some level of noise) is an ill-posed problem (Tikhonov and Arsenin, 1977); thus, there are multiple solutions compatible with the data, and those solutions may not be continuous with respect to data variations (so, small perturbations in the data may cause large differences in the final solutions, preventing self-consistent outcomes). Actually, regularization is about introducing in the inversion process the prior information about the kind of solutions we expect. Hence, regularization selects, amongst all the possible solutions compatible with the data, the unique one that is also in agreement with our expectations (Zhdanov, 2015, 2006). So, in inverting our datasets, we tested different kinds of regularizations; e.g., smooth, sharp, spatially constrained, laterally constrained, and each with different settings. Strictly speaking, since those diverse results fit the data at the same level, they were equally good from a purely geophysical perspective; thus, the intervention of the geologists, with their overall understanding of the possible geological structures and expectations, was crucial. From an epistemological perspective, we can say that, to some extent, the geophysics have been used to falsify some of the geological alternatives, i.e., those that also did not fit the geophysical data (Tarantola, 2006). The models in agreement with both the geophysical data and the geological expectation have been used to (qualitatively) assess the uncertainty of the geophysical models (Fig. 3). In this respect, stochastic inversion would have been the optimal solution to explore (quantitatively) the ambiguities of the geophysical solutions; unfortunately, there are no stochastic inversion schemes we could practically use for problems at this scale. On top of this, propagating the uncertainty of the geophysical solutions into the uncertainty of the geo-model is still an open (and extremely relevant) issue.

For the large majority of the Nasia sub-catchment, a smooth regularization has been used with extremely loose (compared with those generally used for the standard dB–dt data inversion; e.g., Viezzoli et al., 2010, 2013) lateral and vertical constraints. The result is a quasi-3D resistivity volume generated from B-field GEOTEM data that is significantly different (in terms of possible geological interpretation) from the original CDIs (Fig. 4).

### 2.3 3D geo-modeling procedure

Raw data collected from various sources can be interpreted in terms of spatial variations (providing information about the geometries) and/or in terms of the absolute value of the attribute retrieved from the data (characterizing not only the geometry of the features, but also their nature).

The spatial information from the geophysics was used to create a 3D geometry model. Geometric modeling involves two steps. The first concerns the development of a suitable geometric representation of the fundamental geological "framework"; the second relates to the discretization of this framework to provide control for the analytical computations within the numerical models used in the predictive modeling (Turner, 2006).

In the present research, the first stage in the geometric modeling involved the interpolation of inverted 1D AEM data (Fig. 5) into a 3D grid with an assigned search radius of 20 km and a cell size of 2 km for the regional data as well as a search radius of 500 m and a cell size of 100 m for the dense area. The assigned search radius should not be less than the spacing between flight lines to obtain a continuous electrical resistivity distribution (Pryet et al., 2011), but, at the same time, it needs to be small enough to prevent the smearing of possibly useful information. The second, more laborious, step involved constructing the surfaces that define the overall units. Here, both the AEM and borehole data were correlated with a particular stratigraphic unit, and the boundaries for that unit were drawn. This is necessary, and very tricky, since the electrical resistivity as it is inferred from the AEM cannot be unambiguously made to correspond to a specific lithology and/or stratigraphic unit. For this task, knowledge of the geo-physical response behavior and the experience of geologists in outlining which signatures belong to which stratigraphic unit are clearly equally crucial. For instance, low resistivity signatures within the Bombouaka may belong to the Poubogou formation, whereas anomalies with similar low resistivity ranges within the Oti may belong to the Bimbila forma-tion. Also for this reason, the tight interaction between geologists and geophysicists (through several iterations) has been found crucial for effective geo-modeling – for example, in interpreting geological features wherein the geophysical reliability is reduced as we get closer to the depth of investigation (DOI – the shaded portion at the bottom of Fig. 4b).

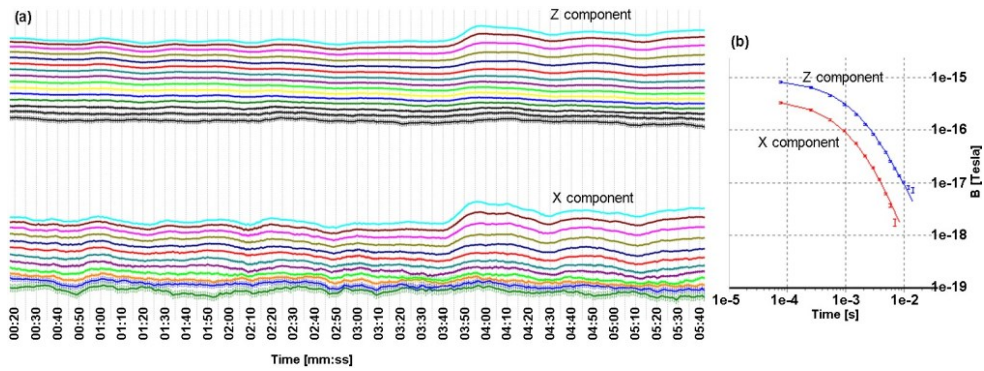

**Figure 2. (a)** An example of the Z and X components of the B-field data obtained after the application of the moving window stacking with width varying with the time gates. **(b)** Example of a typical sounding (Z and X components): the vertical bars represent the stacked data with the associated uncertainty; the solid lines are the calculated data corresponding to the inversion model (not shown).

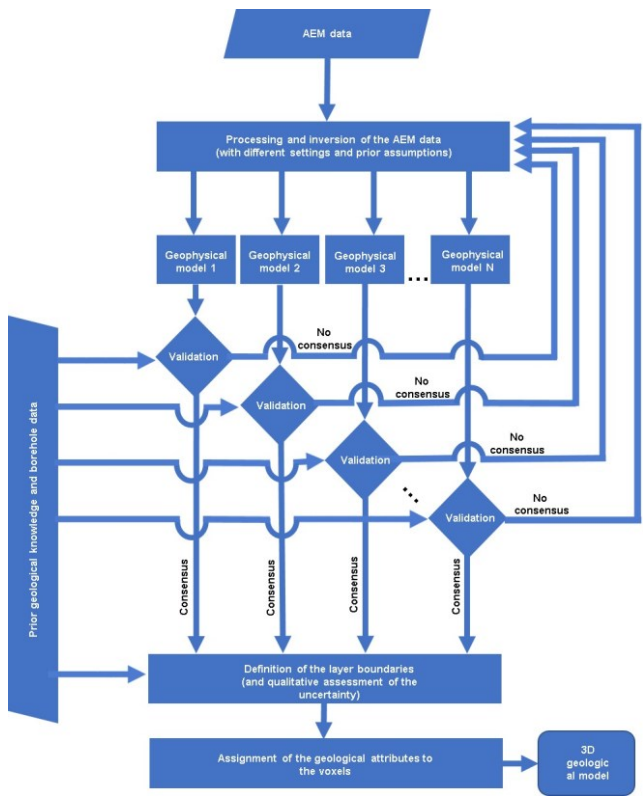

**Figure 3.** The workflow describing the iterative interaction between geologists and geophysicists leading to the development of the 3D geological model consistently integrating all the diverse pieces of information available (geophysical data, prior geological knowledge, wells, etc.).

Thus, the geo-modeling can be considered a way to compile, in a consistent manner, geological knowledge about the area, information from the dense geophysics (which acts as a "smart" interpolator between the available boreholes), and the other available data. In this respect, it is worthwhile to note that only boreholes that had a distance smaller than 1–3 km from the regional flight lines were considered sufficiently representative for the geophysical interpretation.

The outlined boundaries were then used in the next stage for populating the model grid (Ross et al., 2005; Sapia et al., 2015; Jørgensen et al., 2013). Populating the model grid is done by adding and editing voxel groups based on a cognitive approach (Fig. 6; Høyer, et al., 2017, 2015; Jørgensen et al., 2013).

## 3 Results and discussions

### 3.1 Inverted AEM data

Figure 4 shows an example of the comparisons between the original CDIs and the new inversion obtained with the discussed smooth SCI approach applied on the B-field GEOTEM data. The differences are evident. Not surprisingly, the CDI result is characterized by higher lateral variability, as each sounding is converted into a resistivity profile independently, while the SCI, by definition, enforces some degree of spatial coherence. The more prominent CDI's lateral heterogeneity is clear not only on the NE side of Fig. 4a in the shallow portion of the section, where distinct resistive inclusions are detected, but also at depth along all the flight lines where there are spurious lateral oscillations of the elec-trical properties. The SCI result is laterally more consistent; however, this does not prevent the reconstruction of a resistive body (associated with the hotter colors), at a distance of approximately 10 km (circled in red – Fig. 4b) that is well-separated from the resistive superficial unit – continuing on the right – by a clear conductive formation (very differently from what is retrieved by the CDI). In addition, the SCI result shows interesting resistive features incised into the more conductive surroundings (in particular, see the two deepening structures located between 20 and 30 km – circled in black in Fig. 4b). The considerable depth of investigation (DOI – indicated as a white mask in Fig. 4b) is worth noting; generally, the geophysical model parameters can be considered sensitive to the data down to the considerable depth of ∼ 500 m.

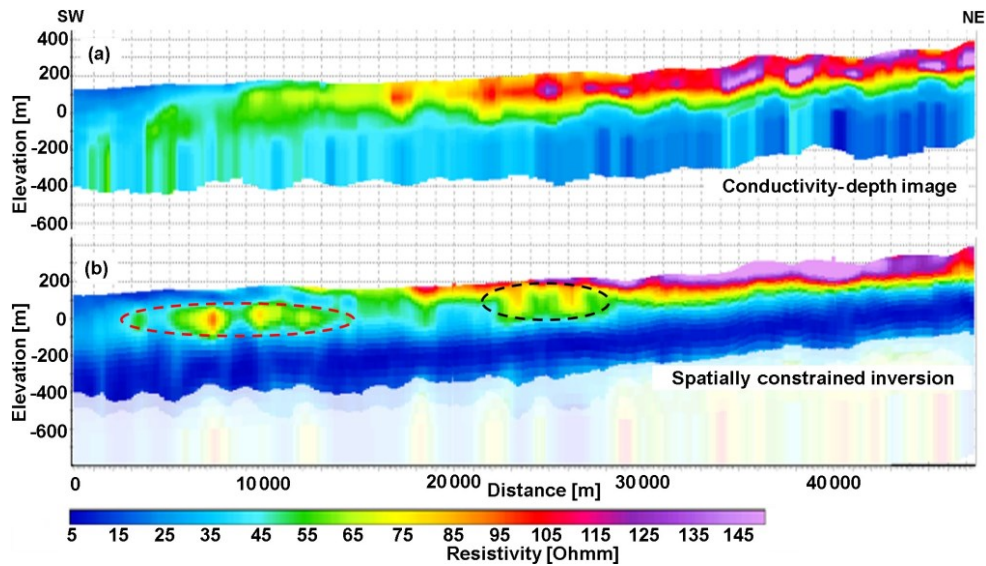

**Figure 4.** **(a)** The original conductivity–depth image (CDI) in a portion of NE–SW line 2 (Fig. 1a); **(b)** the associated results obtained with the new data processing and inversion approach (smooth spatially constrained inversion).

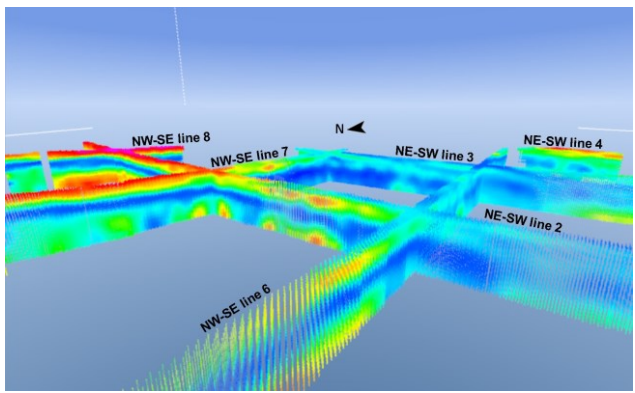

**Figure 5.** A 3D view of the B-field SCI results along the 20 km × 20 km grid lines in the study area. These soundings were used as a basis for geologic interpretation and modeling. The arrowhead points northwards. For the locations of the grid lines, see Fig. 1a.

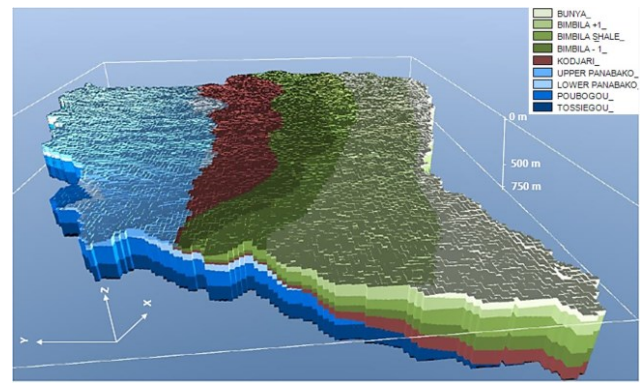

**Figure 6.** 3D geological model of the Nasia sub-basin resulting from the combined interpretation of the B-field airborne data, the prior geological knowledge of the area, and the available wells.

This not only demonstrates the quality of the original data, but also confirms that the survey was designed for deep exploration and not for high-resolution shallow investigations. Therefore, the new SCI provides important insights on the geological settings and highlights resistive, relatively shallow structures, possibly relevant as groundwater reservoirs.

In order to proceed further with the geological interpretation of the geophysical model, the SCI result was gridded (Fig. 5). The general signature trends visible in such a resistivity grid can be summarized as follows:

– areas with low resistivity values characterize argillaceous layers in both the Bombouaka and Oti groups;

– sandstones have characteristically high resistivity values, with the massive quartzose sandstones of the Bom-

bouaka group (specifically, the Panabako sandstone, Anyaboni sandstone, and upper Damongo formation) displaying the lowest conductivities (Fugro Airborne Surveys Interpretation, 2009b).

Figure 6 shows the 3D geological model of the study area. It is mainly based on the geophysics and the other source of available information (regolith outlines from previous radiometric survey; Geological Survey Department, 2006). The developed geological model (consisting of 17.5 million 500 m $\times$ 500 m $\times$ 5 m voxels) shows a coherent 3D representation of the subsurface within the Nasia Basin; it generally honors the available geologic knowledge as well as the information from the wells and the AEM evidence. At the same time, it provides some new insights into the geology of the terrain.

### 3.2 Lithostratigraphy from AEM

Figure 6 shows nine distinct stratigraphic units in the study area. These include the following: (i) Bunya (youngest), (ii) Bimbila $+1$, (iii) Bimbila shale, (iv) Bimbila $-1$, (v) Kodjari formation, (vi) upper Panabako sandstone, (vii) lower Panabako sandstone, (viii) Poubogou formation, and (ix) Tossiegou formation (oldest).

#### 3.2.1 The Bombouaka group

In the study area, the Poubogou and Panabako formations of the Bombouaka group outcrop in the north. In contrast, outcrops of the basal unit of this group, the Tossiegou formation, have not been observed within the study area.

**Tossiegou formation**

This is the oldest unit of the Bombouaka group identified by resistivity signatures ranging between approximately 30 and 120 $\Omega$m on cross sections of the inverted AEM volume. The formation is comprised of basal argillaceous strata that grade upwards into feldspathic and quartzitic sandstones (Carney et al., 2010). They overlie crystalline basement rocks of the Birimian (Anani et al., 2017). For example, from the NE–SW section across the resistivity volume shown in Fig. 7, the Tossiegou formation is seen to extend way beyond 140 m below sea level. An estimation of the thickness of the formation is, however, made difficult by its extension below the DOI. In fact, the great depth of the formation, together with the overlying more conductive layers (the Bimbila and Poubogou formations), generally prevents the electromagnetic signal from propagating to greater depths, making the inferred resistivity values of that formation less sensitive to the data (and therefore difficult to precisely resolve).

**Poubogou formation**

This unit is identified within the Gambaga escarpment with an average thickness of 170 m (Fig. 7). The basin-wide distribution of this sequence indicates a possible regional transgression event (Fugro Airborne Surveys Limited, 2009b). The formation consists of green–grey micaceous mudstones and siltstones intercalated with sandstones at some places. As it grades into the overlying Panabako formation, there is an increase in the sandstone proportion relative to the argillaceous beds (Carney et al., 2010). This formation exhibits low resistivity in AEM profiles ranging between 0 and 20 $\Omega$m and appears to have a thickness in the range of 150–180 m along a NE–SW profile in the study area. This is consistent with thicknesses recorded by Carney et al. (2010).

**Panabako formation**

This is a quartz-arenite-rich formation with a suggested thickness of 150–200 m (Carney et al., 2010). Lithostratigraphic mapping by Ayite et al. (2008) identified two subdivisions of the Panabako formation within the Nakpanduri escarpment. The upper division consists of a nearshore aeolian sequence, while the lower sequence is composed of a nearshore fluvial sequence (lower Nakpanduri sandstone formation); Carney et al. (2010) correlate the lower Nakpanduri with the upper Poubogou. From the current AEM data, this subdivision is, however, observed entirely within the Panabako with the presence of a distinct resistivity contrast clearly visible in the newly inverted data (e.g., Fig. 7); indeed, in the new AEM reconstruction, the upper Panabako shows higher resistivities – ranging from approximately 60 to 200 $\Omega$m – and the lower layer is characterized by relatively moderate conductivity – roughly between 30 and 60 $\Omega$m. The tendency of the Bombouaka group sandstone units to fine towards argillaceous strata at their base (Jordan et al., 2009) can be inferred from the increasing conductivity values towards the base, indicating a probable increase in argillaceous material.

#### 3.2.2 The Oti group

This group underlies the southern portion of the study area. Generally, it records the transition from a shallow marine environment adjacent to a rifted margin into a marine foreland basin sequence represented by interbedded argillaceous and immature arenaceous material (Carney et al., 2010).

**Kodjari formation**

Composed of what is commonly known as the triad, this formation constitutes the basal unit of the Oti group (Fig. 6). Commonly, the Kodjari formation comprises (i) basal tillites, followed by (ii) a cap-carbonate limestone, and finally covered with (iii) laminated tuffs and ash-rich siltstones (Carney et al., 2010).

The presence of the Kodjari formation is not easily seen but can be inferred from the SCI resistivity sections by moderately resistive strata observed immediately above the topmost units of the Bombouaka group (Fig. 7). An average thickness of 75 m can be retrieved; however, it should be noted that its continuity throughout the basin has not been verified. Carney et al. (2010) noted that, at some localities, in the north of the Volta Basin, the overlying tuffaceous material of the Kodjari triad is seen to unconformably lie directly on the Panabako rocks of the Bombouaka as a result of the lateral discontinuity of these units. These occurrences are also confirmed in the reprocessed AEM data (e.g., Fig. 7, at around 40–45 km).

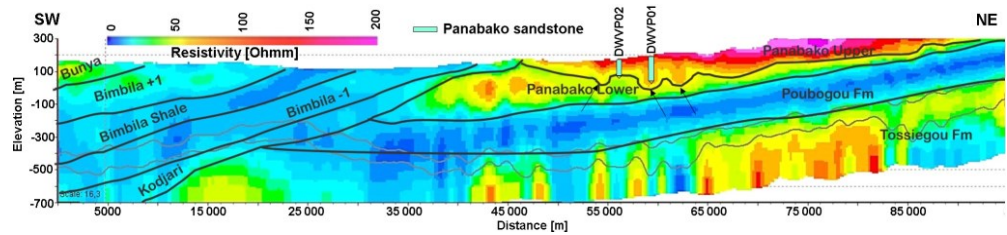

**Figure 7.** Cross section along SW–NE line 2 across the study area (see Fig. 1a for the location) with the conceptual geological interpretations showing the U-shaped valleys (between 53 and 63 km, whose location is indicated by three black arrows). In addition, two of the geologic logs (DWVP02 and DWVP01; Fig. 1a) used for the demarcation of lithostratigraphic boundaries and the interpretation and/or verification of the geophysical model are also shown. The two solid grey lines at the bottom represent the DOIs.

**Bimbila formation**

The Bimbila formation has two sandstone beds forming its upper and lower boundaries. These are the Chereponi sandstone member, which forms the basal stratum of the formation, and the Bunya sandstone member, which generates the exposed upper portion of the formation. The Bunya sandstone is observed as a moderately conductive layer in the AEM cross section, above the argillaceous material of the Bimbila (Fig. 7).

The argillaceous units of the formation consist mainly of green to khaki micaceous laminated mudstones, siltstones, and sandstones representing a continuation of foreland basin deposition.

### 3.3 Structural interpretations

The new results from the inversion of the AEM data reveal some amount of deformation within the basin. Dips of approximately 20° to the SW of the Bimbila are seen from AEM interpretations, giving an indication of the arcuate nature of the basin.

Along NE–SW line 7 (Fig. 8), a vertical displacement is observed and is interpreted as a fault within the Bimbila. It aligns well (sensu lato) with late brittle faults (Crowe and Jackson-Hicks, 2008).

An angular unconformity marking the transition between rocks of the Bombouaka (which began, according to Carney et al., 2010, to accumulate after 1000 Ma) and the Oti rocks (which have a maximum depositional age of 635 Ma; Carney et al., 2010) is observed in Fig. 7. The unconformity could possibly be related to the absence of zircons aged 950–600 Ma recorded by Kalsbeek et al. (2008), suggesting the presence of an oceanic gap that prevented the deposition of sediments. The unconformity separates continental deposits of the Bombouaka below from passive margin deposits of the Oti above (Kalsbeek et al., 2008).

**Paleovalleys**

Three characteristic U-shaped valleys towards the north of the basin, which are between 53 and 63 km along the profile in Fig. 7 (highlighted by black arrows), were interpreted from AEM data to be resistive features at the base of the upper Panabako cutting into the lower Panabako. The valleys exhibit a NW–SE trend (Fig. 9) and are considered to be tunnel valleys (Jørgensen and Sandersen, 2006; Kehew et al., 2012; Van der Vegt et al., 2012) whose origin is still to be fully investigated. The presence of these valleys may be of stratigraphic interest as well as hydrogeologic significance.

The proposed presence of valleys between the upper and lower Panabako sequences represents an unconformity before the deposition of the upper Panabako sequence (Fig. 7). The geometry of the valleys, with their U-shaped cross-sectional nature, leads to the deduction that glaciation could have played a role in their formation. Moreover, new insights into the stratigraphy may be implied by the possible presence of these valleys within the Panabako formation. The high-energy event responsible for producing the intra-formational unconformity most likely occurred within the wide age range of 1000 to 635 Ma (Carney et al., 2010). The upper limit is defined by the detrital zircon analysis of the Bombouaka group (Kalsbeek et al., 2008), whereas the lower limits, representing the period of deposition of the tillites and diamictons of the Oti group within the Kodjari formation, should correspond to the end of the Cryogenian glacial period, which has been dated at 635 Ma by Carney et al. (2010). This possible event would, however, be younger than the 1100 Ma deposition of sediments of the Bombouaka based on the detrital zircon analysis discussed in Kalsbeek et al. (2008). These valleys, then, represent a distinct history of glaciation separate from the Marinoan glaciation recorded in the Kodjari (Porter et al., 2004). This is similar to what is seen in the Wassangara group (Deynoux et al., 2006; Shields-Zhou et al., 2011) of the Taoudéni Basin outcropping in western Mali and southern Mauritania. Located in the southern region of the Taoudéni Basin, thick successions of glacial influence have been recorded (Shields-Zhou et al., 2011) and were initially thought to form part of Supergroup 2 of the Taoudéni Basin. However, the paleovalley presence below the craton-wide erosional and angular unconformity marking the transition between Supergroups 1 and 2 precludes their association with the Marinoan glaciation of Supergroup 2 and includes them in what is referred to as the Wassangara group rocks of Supergroup 1 (Deynoux et al., 2006; Shields-Zhou et al., 2011). Previous research within the Voltaian is replete with information on the glaciation within the Neoproterozoic Marinoan, based on which an unconformity between the top of the Bombouaka group and the basal units of the Oti–Pendjari group was proposed. This unconformity is conspicuously marked by the triad, consisting of basal tillites, cap carbonates, and silicified tuffs (Goddéris et al., 2003).

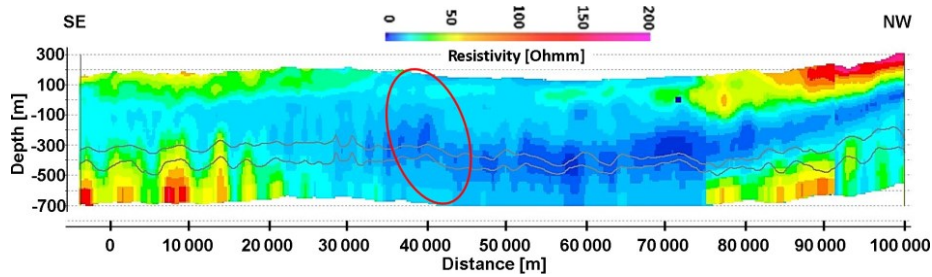

**Figure 8.** Cross section along NW–SE line 7 (Fig. 1a) showing faulting (within the red circle) in the Bimbila. The DOI is shown as a solid grey line (there are two of them according to their definition – more details on this distinction can be found in Christiansen and Auken, 2012).

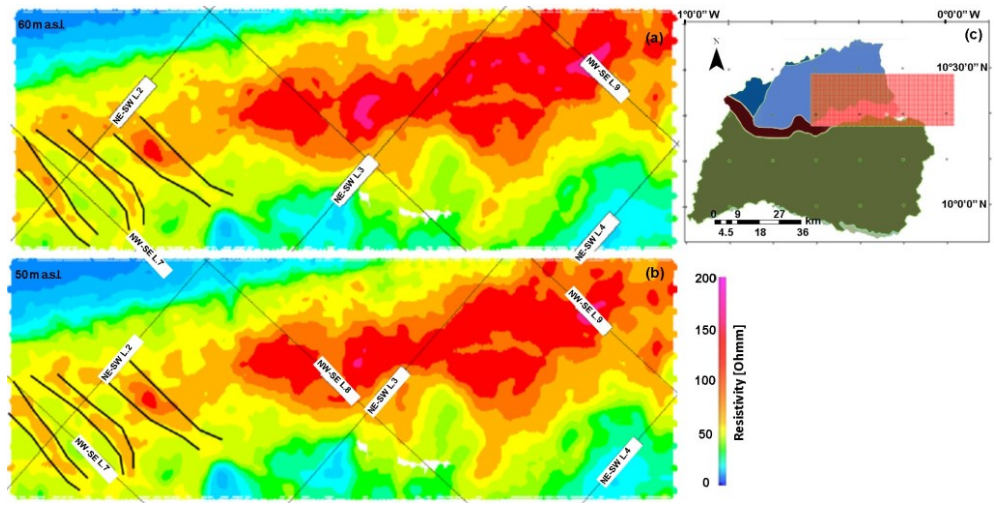

**Figure 9.** Horizontal resistivity slices at 60 m **(a)** and 50 m **(b)** of depth in the portion of the study area – in red, panel **(c)** – characterized by a geophysical sampling much denser (200 m line spacing) than the regional survey – in this respect, see the lines from NE–SW L.2 to NW–SE L.9 in panels **(a, b)** characterized by a 20 km × 20 km spacing (Fig. 1a). The black solid lines in the bottom left corner of panels **(a, b)** show the location of the paleovalleys discussed in the paper. Hence, the features interpreted as paleovalleys can be identified in both independently inverted datasets (i.e., the dense coverage area and the regional survey). Clearly, the densely sampled data can provide further insights in terms of the spatial coherency of the paleovalley NW–SE trend.

On the other hand, Deynoux et al. (2006) mention that the 400–500 m thick glacially influenced succession was controlled by the tectonic evolution of the nearby Pan-African belt with deposition at around 660 Ma. These proposed pre-Marinoan, or possibly Sturtian (∼717 to 643 Ma), glacial events and deposits are suggested to be related to mountain glaciers (Bechstädt et al., 2018; Deynoux et al., 2006; Hoffman and Li, 2009; Villeneuve and Cornée, 1994). Though these assertions are hypothetical, they might be consistent with the high paleolatitude of the West African craton during the Proterozoic (Bechstädt et al., 2018; Hoffman and Li, 2009) combined with other geophysical and correlative evidence on the same craton (Dzikunoo et al., 2018; Shields-Zhou et al., 2011).

The rocks of the Bombouaka group in the Voltaian sedimentary basin are said to be reminiscent of the rocks in portions of Supergroup 1 in the Taoudéni Basin (Shields-Zhou et al., 2011), and the presence of glacial signatures in both groups suggests that the pre-Marinoan glaciation must have been regional. The trends of the paleovalleys in the study area, i.e., NW–SE (Fig. 9), align well with paleogeographic reconstructions of glaciation in the NW Africa region, which suggests the presence of an ice sheet towards the north of the Reguibat shield with inferred glacial movement southwards towards the Pan-African belt (Shields-Zhou et al., 2011). The glacial movement is further verified by the transition of sediments in the region from glacial to a mixture of glacial and marine and finally marine towards the border with the rocks of the Pan-African belt. Some authors consider the combination of Sturtian and Marinoan glaciations – both of the Cryogenian – to suggest a complete glaciation event; i.e., the Snowball Earth, according to which both continental and oceanic surfaces were covered by ice (Goddéris et al., 2003; Hoffman and Li, 2009; MacGabhann, 2005). A possible point of contention for the Snowball Earth hypothesis – that clearly suggests the ubiquitous presence of ice sheets during the Marinoan and Sturtian (including the warmest parts proximal to the paleo Equator as well as the higher paleolatitudes and high elevations) – is the lack of such glacial evidence in the West African craton from the Sturtian. In fact, within the framework of that hypothesis, quite complex assumptions (Hoffman and Li, 2009) have been made in an attempt to justify the absence of glacial traces in the West African craton located, during the Sturtian, in cold regions (latitude around 60° S; Lie et al., 2008). So, the presence of the discussed glacial paleovalleys in the Voltaian can easily fill the gaps with no need for additional ad hoc assumptions. On the other hand, the lack of pervasive evidence of the Sturtian glaciation within the Voltaian and the Taoudéni could be due to the overprinting of glacial structures by the more recent Marinoan glaciation or tectonic activity related to regional subsidence during the evolution of the Voltaian (Ayite et al., 2008). Outcrop investigations of samples from the Bombouaka, however, do not show mixtites–diamictites, which are typical of glacial deposits and could suggest a reworking of the glacial deposits by some fluvial action; this seems quite similar to the situation Bechstädt et al. (2018) refer to as post-glacial transgression, resulting in the infilling of incised valleys with fluvial, reworked glacial and marine deposits. Carney et al. (2010) observed two sandstone sequences in the Panabako, with the upper unit forming "sugarloaf" cappings above the lower sandstones. These structures are characteristic of high-energy environments (Ayite et al., 2008) and may also be remnants of the Sturtian glaciation reworked by some marine or fluvial activities.

To summarize, the geological interpretation of the newly reprocessed AEM data (with their significantly enhanced information content) facilitated the discovery of evidence showing the presence of potential paleovalleys, possibly acting as groundwater reservoirs. At the same time, the existence of such geological features and, in particular, their stratigraphic location within the bounds of the Panabako formation suggest the need for a possible revision of the stratigraphy of the Bombouaka group, especially within the study area. Furthermore, the new insights suggest that there was some pre-Marinoan glacial activity responsible for the paleovalleys within the Panabako (Hoffman and Li, 2009) in the Voltaian sedimentary basin; thus, this activity precedes the Marinoan glaciation episode that is generally associated with the Kodjari formation of the basin (Deynoux et al., 2006) but still occurs within the Cryogenian period. This new glaciation suggests the possibility of a Sturtian event, but this assertion is currently hypothetical and would need further investigations to verify. Possible glacial incisions within the Panabako seem reasonable because of the high paleolatitude of the West African craton but, at the same time, avoid the need for additional complex justifications for the absence of indications of ice sheets in poleward continents. Hence, the proposed combination of the Marinoan and Sturtian events in the Neoproterozoic Voltaian sedimentary basin, if verified, would be compatible with the hypothesis of a global Neoproterozoic Snowball Earth even at high paleolatitudes (Hoffman and Schrag, 2002).

### 3.4 Hydrogeological applications of the geological model

The 3D geological model developed in this research is to be used as the basis for conceptualizing the hydrogeological context of the basin and the larger Voltaian supergroup. For instance, the apparent detection of valleys within the Panabako formation may provide an indication of a deeper, prolific aquifer system that has not been noted before in the hydrogeology of the Voltaian supergroup. The presence of such systems in the Voltaian would have significant implications for the large-scale development of groundwater resources for irrigation and other income-generating ventures in the area. The Voltaian supergroup has been noted as a difficult terrain in terms of groundwater resources development, and the Nasia Basin, in particular, is one of the basins where high borehole failure rates have led to chronic domestic water access challenges over several years. Within or after the current DANIDA project, the paleovalleys need to be further investigated, which will lead to both seismic surveys and the drilling of much deeper boreholes penetrating them.

## 4 Conclusions

The present research investigates the concrete possibility of using preexisting airborne electromagnetic data, originally collected for mineral exploration, to build accurate 3D geological models for hydrogeological purposes. The use of this specific kind of data (B-field time-domain electromagnetic measurements) for this scope is quite novel per se and, in this specific case, allowed for the reconstruction of the stratigraphy of the Nasia Basin within the Voltaian sedimentary basin. In particular, the proposed geo-modeling strategy made it possible to infer the presence of paleovalleys that have been identified as pre-Marinoan and may be products of a glaciation event within the Sturtian (old Cryogenian). The valleys correlate with glacial deposits observed in the Wassangara group of the Taoudéni Basin. This group is found within Supergroup 1, which correlates with the Bombouaka rocks of the Voltaian basin. If confirmed, the stratigraphic location of these potential paleovalleys within the Panabako formation would lead to a possible revision of the stratigraphy of the Bombouaka group, especially within the study area. Moreover, together, the paleovalleys and the glacial deposits give further evidence for a Snowball Earth event that possibly covered the entire Earth during the Neoproterozoic. So, the impact of these findings goes beyond the discovery of potential groundwater reservoirs (which by itself is extremely relevant from a socioeconomic perspective) and can contribute to a rethinking of the stratigraphy of the region and confirm the Neoproterozoic Snowball Earth hypothesis.

*Data availability.* The rights to the data used in this research are owned by Geological Survey Authority of Ghana. The authors therefore do not have the right to make the data available to the public.

*Author contributions.* EAD contributed to investigation, data curation, methodology, visualization, writing the original draft, and review and editing. GV contributed to conceptualization, funding acquisition, investigation, data curation, methodology, software and algorithm development, supervision, validation, and review and editing. FJ contributed to investigation, methodology, supervision, validation, and review and editing. SMY contributed to conceptualization, funding acquisition, project administration, supervision, validation, and review and editing. BBY provided supervision.

*Competing interests.* The authors declare that they have no conflict of interest.

*Acknowledgements.* The authors would like to thank DANIDA for its support of this research through the South-driven project "Ground Water Development and Sustainable Agriculture (proj. code: 14-P02-GHA)", also known as "GhanAqua", and the Geological Survey Authority of Ghana for providing most of the data and for its invaluable help. In this respect, special thanks go to Emmanuel Mensah and the director, John Agyei Duodu. In addition, the authors are very grateful to Kurt Klitten and Per Kalvig from the Geological Survey of Denmark and Greenland for their love of Ghana and for making this adventure possible.

*Financial support.* This research has been supported by the DANIDA Fellowship Centre (grant no. 14-P02-GHA).

*Review statement.* This paper was edited by Ulrike Werban and reviewed by Richard Smith, Giorgio Ghiglieri, and one anonymous referee.